\documentclass[aps, prl, reprint,notitlepage,superscriptaddress,showpacs]{revtex4-1}
\usepackage{amssymb}  
\usepackage{amsfonts}  
\usepackage{amsmath}  
\usepackage{amsthm}  
\usepackage{bm}  
\usepackage{bbm}  
\usepackage[colorlinks=true]{hyperref}
\usepackage{comment}
\usepackage{graphicx}
\usepackage{tabularx}
\usepackage{multirow}
\usepackage{bbding}
\usepackage{pgfplots}

\definecolor{mygrey}{RGB}{225, 225, 225}

\begin{document}

\title{Hidden spin current in doped Mott antiferromagnets}

\author{Wayne Zheng}
\affiliation{Institute for Advanced Study, Tsinghua University, Beijing, 100084, China}
\author{Zheng Zhu}
\affiliation{Department of Physics, Massachusetts Institute of Technology, Cambridge, MA, 02139, USA}
\author{D. N. Sheng}
\affiliation{Department of Physics and Astronomy, California State University, Northridge, CA, 91330, USA}
\author{Zheng-Yu Weng}
\affiliation{Institute for Advanced Study, Tsinghua University, Beijing, 100084, China}
\affiliation{Collaborative Innovation Center of Quantum Matter, Tsinghua University, Beijing, 100084, China}

\date{\today}


\begin{abstract}

We investigate the nature of doped Mott insulators using exact diagonalization and density matrix renormalization group methods.  Persistent spin currents are revealed in the ground state, which are concomitant with a nonzero total momentum or angular momentum associated with the doped hole. The latter determines a nontrivial ground state degeneracy. By further making superpositions of the degenerate ground states with zero  or unidirectional spin currents, we show that different patterns of spatial charge and spin modulations will emerge. Such anomaly persists for the odd numbers of holes, but the spin current, ground state degeneracy, and charge/spin modulations completely
disappear for even numbers of holes, with the two-hole ground state exhibiting a d-wave symmetry. An understanding of the spin current due to a many-body Berry-like phase and its influence on the momentum distribution of the doped holes will be discussed.

\end{abstract}

\maketitle

\emph{Introduction.---}The low-energy physics of the interacting fermions is generally described as a Luttinger liquid (LL) \cite{luttinger1963}\cite{tomonaga1950} in one dimension (1D), characterized by gapless charge, neutral density wave and current excitations \cite{haldane1981}\cite{PhysRevLett.79.1110}. In general, the LL theory breaks down in higher dimensions due to the absence of forbidden regions in the spectrum to protect the current excitations, with the exception for some fractional quantum Hall systems \cite{PhysRevLett.88.036401}\cite{PhysRevB.89.085101} in two dimensions (2D) where the gapless edges are protected by the gapped bulk.  Another class of strongly interacting fermion systems is the doped Mott insulators, relevant to high-temperature superconducting cuprates \cite{anderson1997}\cite{RevModPhys.78.17}\cite{RevModPhys.78.17}, for  which Anderson \cite{anderson1997}\cite{PhysRevLett.64.1839}\cite{PhysRevLett.65.2306} was the first to suggest that doped holes may induce scattering singularities  leading to LL-like behaviors in 2D. Its microscopic mechanism was attributed \cite{PhysRevLett.64.1839} to an \emph{unrenormalizable Fermi-surface phase shift} generated by the doped holes, which was later identified with a many-body Berry-like phase in the $t$-$J$ model known as the \emph{phase string} \cite{PhysRevLett.77.5102}\cite{PhysRevB.55.3894}\cite{PhysRevB.77.155102}. The latter is  responsible for the LL behaviors in the 1D $t$-$J$ model as confirmed both analytically and numerically \cite{PhysRevB.55.3894}\cite{zhu201651}. Then a natural question is if such an effect can lead to a current-carrying ground state \cite{SS1988}\cite{Weng2011} in the 2D doped Mott antiferromagnet to give rise to non-Fermi liquid (NFL) features.

\begin{figure}[]
    \centering
    \includegraphics[width=.45\textwidth]{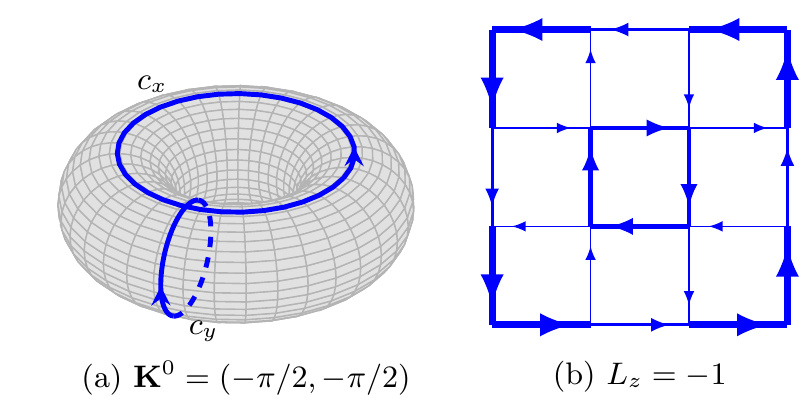}
    \caption{Neutral spin currents are revealed in the degenerate ground states of the one-hole-doped $t$-$J$ model on square lattices: (a) on a torus (PBC) with $c_{x}$ and $c_{y}$ denoting two winding paths at momentum ${\bf K}^0=(-\pi/2, -\pi/2)$ (cf. Table I); (b) the spin current pattern under the OBC (angular momentum $L_z=-1$).  Here $J/t=0.3$ with fixed $S^z=1/2$.     }
    \label{fig:torus}
\end{figure}

In this paper, we reveal unconventional properties of the doped Mott antiferromagnets based on exact diagonalization (ED) and density matrix renormalization group (DMRG) simulations. For the odd numbers of doped holes, we identify the symmetry-protected degeneracy with nontrivial total momentum ${\bf K}^0\neq0$ or angular momentum $L_z\neq 0$ for the ground states, and more importantly,  it is concomitant with permanent spin currents, as illustrated in Fig. \ref{fig:torus} by taking one-hole ground state as an example. Such spin current pattern is robustly present in different sample sizes and parameter regimes, 
adapting to different geometries [e.g., under the periodic boundary condition (PBC) in Fig. \ref{fig:torus}(a) and under open boundary condition (OBC) in Fig. \ref{fig:torus}(b)]. It indicates a nontrivial many-body Berry-like phase induced by the doped holes. In particular, by making superpositions of  the degenerate ground states with diminished or unidirectional spin currents, we show that different patterns of the spatial charge and spin modulations emerge. In contrast, the degeneracy and its associated spin currents disappear simultaneously for the even numbers of holes, say, in the two-hole ground state, which exhibits a d-wave symmetry. Such even-odd effect persists over a few hole cases and may have important implications for finite doping, which is potentially relevant to the superconductivity and pseudogap physics in high-$T_c$ cuprates~\cite{anderson1997}\cite{RevModPhys.78.17}\cite{nature14165}.

\begin{table}[!h]
    \caption{Momenta and spin currents of degenerate one-hole ground states on a $4\times4$ torus determined by ED.}
    \centering
    \begin{ruledtabular}
        \begin{tabular}{p{.04\textwidth}p{.13\textwidth}p{.15\textwidth}p{.16\textwidth}}
            $J/t$ & $(K_{x}^{0}, K_{y}^{0})$ & $I_{s}^{x}\equiv \sum_{c_x}J^s_{ij}$ & $I_{s}^{y}\equiv \sum_{c_y}J^s_{ij}$ \\
            \hline
            \multirow{6}{*}{$0.3$} & $(0, \pi)$ & $0.0000$ & $0.0000$ \\
            & $(\pi, 0)$ & $0.0000$ & $0.0000$ \\
            & $(\pi/2, \pi/2)$ & $-0.0991$ & $-0.0991$ \\
            & $(\pi/2, -\pi/2)$ & $-0.0991$ & $+0.0991$ \\
            & $(-\pi/2, -\pi/2)$ & $+0.0991$ & $+0.0991$ \\
            & $(-\pi/2, \pi/2)$ & $+0.0991$ & $-0.0991$ \\
            \hline
            \multirow{4}{*}{$3.0$} & $(\pi/2, 0)$ & $-0.0359$ & $0.0000$ \\
            & $(-\pi/2, 0)$ & $+0.0359$ & $0.0000$ \\
            & $(0, \pi/2)$ & $0.0000$ & $-0.0359$ \\
            & $(0, -\pi/2)$ & $0.0000$ & $+0.0359$ \\
            \hline
            {$10$} & $(0, 0)$ & $0.0000$ & $0.0000$ \\
        \end{tabular}
    \end{ruledtabular}
    \label{tab:degenerate_gs}
\end{table}

We shall study the simplest model of a doped Mott insulator, i.e., the $t$-$J$ model, which reads
\begin{equation}\label{tj}
    \begin{split}
        H_{t} &= -t\sum_{\langle{ij}\rangle, \sigma}(c_{i\sigma}^{\dagger}c_{j\sigma}+h.c.), \\
        H_{J} &= J\sum_{\langle{ij}\rangle}\left(\mathbf{S}_{i}\cdot\mathbf{S}_{j}-\frac{1}{4}n_{i}n_{j}\right).
    \end{split}
\end{equation}
Here,  ${{c_{i\sigma }^{\dagger }}}$ is the electron creation operator at site $i$, ${\mathbf{S}_{i}}$ the spin operator, and ${n_{i}}$ the electron number operator, and the summation is over all the nearest-neighbor (NN) sites $\langle ij\rangle$. The Hilbert space is always constrained by the no-double-occupancy condition, i.e., $n_{i}\leq 1$.
We use both ED \cite{arpackpp} and DMRG \cite{PhysRevLett.69.2863}\cite{RevModPhys.77.259} to study the ground states of Eq.~(\ref{tj}) on a 2D lattice of size
$N=N_x\times N_y$.

\emph{Ground state degeneracy and hidden spin currents.---}We begin with the one-hole case, whose basic  properties have been previously intensively investigated \cite{RevModPhys.66.763}\cite{PhysRevB.40.9035}\cite{PhysRevB.47.5984}\cite{PhysRevB.52.R15711} by ED. The ground state has a total spin $S=1/2$ and nonzero momentum (or angular momentum) depending on the ratio $J/t$ for a fixed spin $\hat{z}$-component $S^z=\pm1/2$. For example, for $N=4\times 4$ and $N=12 \times 4$ systems, ED and DMRG calculations show that the ground states have finite total momenta ${\bf K}^0=(\pm \pi/2, \pm \pi/2)$ at $t/J>1$ with four fold degeneracy~\cite{note1}. TABLE \ref{tab:degenerate_gs} shows the details for the $N=4 \times 4$ lattice under the PBC. In contrast, for a bipartite lattice under the OBC with $\pi/2$ rotational symmetry, a double degeneracy can be generally identified as characterized by angular momentum $L_z=\pm 1$ \cite{note2}, with the sample size persisting from a $2\times 2$ plaquette \cite{HYao2007} up to $8\times 8$ (see below).

Here the degenerate ground states associated with ${\bf K}^0\neq0$ or $L_z\neq 0$ imply that the doped hole acquires a non-dissipative charge current flowing through a neutral spin background. One may further check  the \emph{neutral} spin current in the spin background, defined by
\begin{equation}   \label{sc}
    J^{s}_{ij}\equiv -\text{i}\frac{1}{2}\langle\psi|\left({S}_{i}^{+}{S}_{j}^{-}-S_{i}^{-}{S}_{j}^{+}\right)|\psi\rangle
\end{equation}
on a given NN link $ij$ with the ground state $|\psi\rangle$ labeled by quantum numbers $S$ and $S^z$. Indeed $J^s$ per link is found nonzero as illustrated in Fig. \ref{fig:torus} for both PBC [(a)] and OBC [(b) with the arrow and thickness of each link denoting the current direction and amplitude]. The nontrivial ${\bf K}^0$ at $J/t=0.3$ and $J/t=3.0$ are always associated with non-zero spin currents, $I_s^{x(y)}\equiv \sum_{c_x(c_y)}J^s_{ij}$ (cf. TABLE \ref{tab:degenerate_gs}) along the closed path $c_x$ or $c_y$ defined in Fig. \ref{fig:torus}(a). At $J/t=0.3$ there are actually two more degenerate states at ${\bf K}^0=(\pi, 0)$ and $(0, \pi)$ with vanishing $I_s^{x(y)}$, which may be due to an additional special symmetry for the $4\times 4$ lattice but not generic\cite{PhysRevB.40.9035}.  At $J/t=10$, the nontrivial ground state degeneracy (for each fixed $S^z=\pm 1/2$) and the spin current are both absent, while the total momentum reduces to trivial ${\bf K}^0=(0, 0)$.

\begin{figure}[]
    \centering
    \includegraphics[width=.45\textwidth]{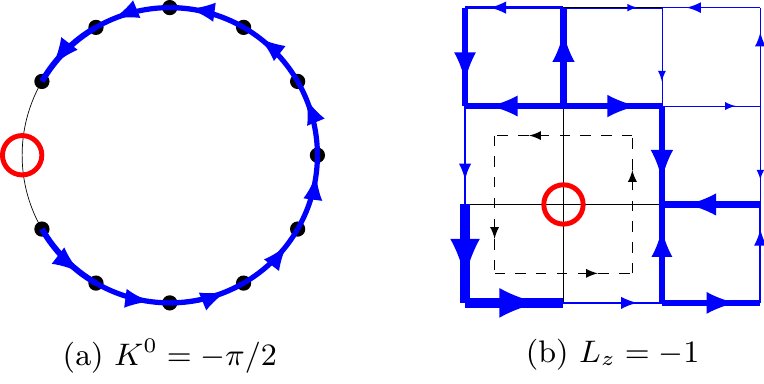}
    \caption{The neutral spin current patterns of $J_{ij}^s$ with a hole projected onto a lattice site at $J/t=0.3$.  (a) For the 1D ground states of a $N=12$ loop; (b) For the 2D ground states of $N=4\times4$ lattice under the OBC. The dashed closed path circulating around the hole indicates a finite net spin current loop. }
    \label{fig:chain}
\end{figure}
Note that $J^s$ in Eq. (\ref{sc}) only satisfies the continuity equation for the conserved $S^z$ at half-filling. Upon doping, to satisfy the full continuity equation, one needs to also include a different
contribution to the spin current at the links involving the hole(s) determined by the hopping term of the $t$-$J$ model, which is also associated with the charge current of the doped hole (cf. the Supplementary Material for details). Nonetheless, $J^s$ in Eq. (\ref{sc}) measures the \emph{neutral} spin current created in the spin background by the hopping term in Eq. (\ref{tj}). To see that such neutral spin current is separated from the hole, we may take the one-dimensional $t$-$J$ chain as an example, in which the one-hole ground state has a double degeneracy at momenta $K^{0}=\pm \pi/2$ (with the lattice size $N=12$ and $J/t=0.3$). By projecting the hole onto a given lattice site, the neutral spin current pattern is shown in Fig. \ref{fig:chain} (a) at $K^{0}=-\pi/2$.
Figure \ref{fig:chain} (b) further shows the neutral spin current pattern with a hole projected onto a specific site in an $N=4\times 4$ lattice under the OBC [cf. Fig. \ref{fig:torus}(b)].

\begin{figure}[]
    \centering
    \includegraphics[width=.4\textwidth]{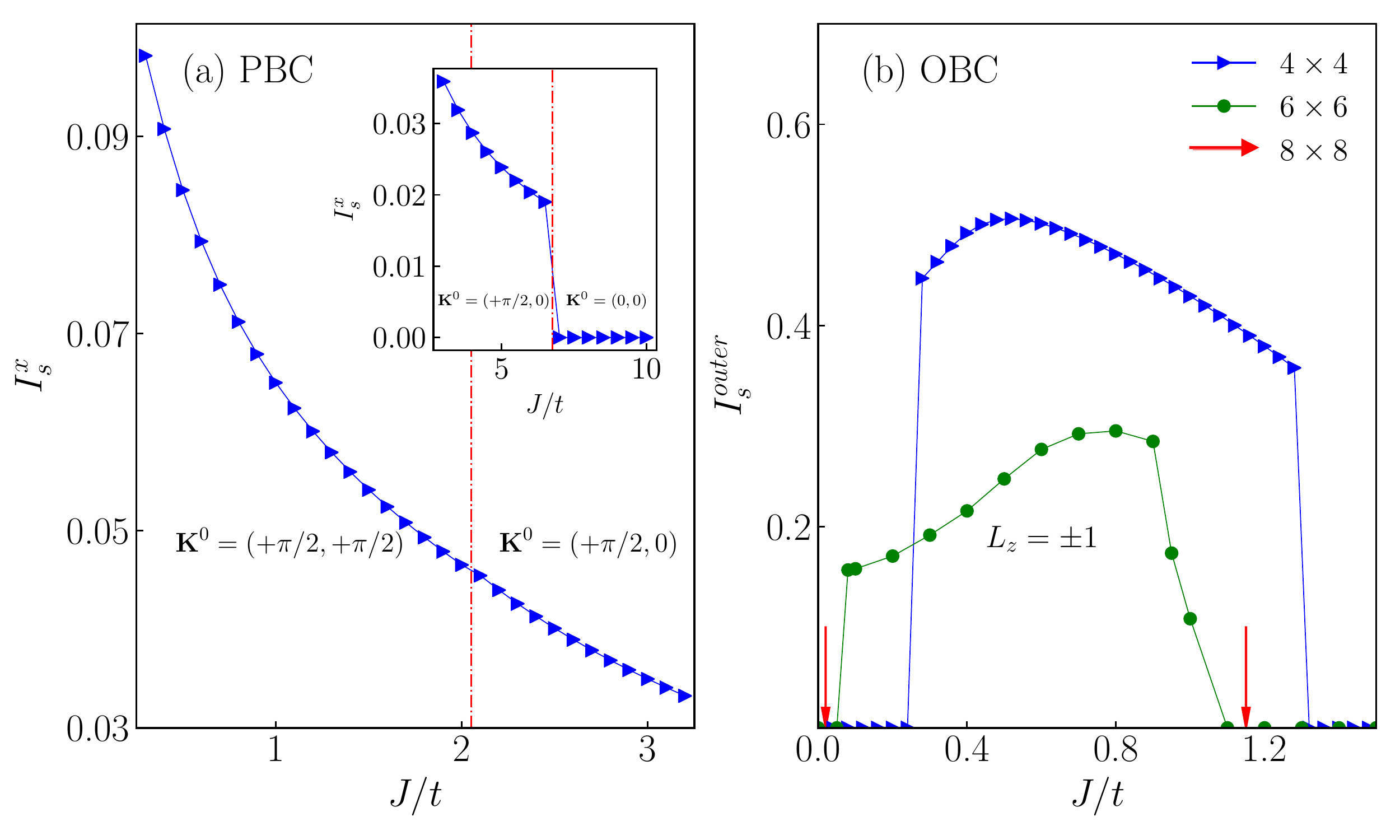}
    \caption{(a) Spin current $I_s^x$ for the single hole ground state of $N=4\times 4$ under PBC. The vertical dashed line marks the jump of the total momentum from $\mathbf{K}^{0}=(+\pi/2, +\pi/2)$ to $\mathbf{K}^{0}=(+\pi/2, 0)$. Inset: the spin current eventually disappears in the non-degenerate ground state with $\mathbf{K}^{0}=0$ at a larger $J/t> 7$; (b) The total spin currents summed over the outermost bonds of the $4\times4$ and $6\times 6$ lattices under OBC, respectively, with the nonzero spin current regimes coinciding with $L_{z}=\pm 1$. The vertical arrows mark the critical points for the $8\times 8$ lattice (see text). }
    \label{fig:spin_current_trend}
\end{figure}

\begin{figure*}[]
    \centering
    \includegraphics[width=.9\textwidth]{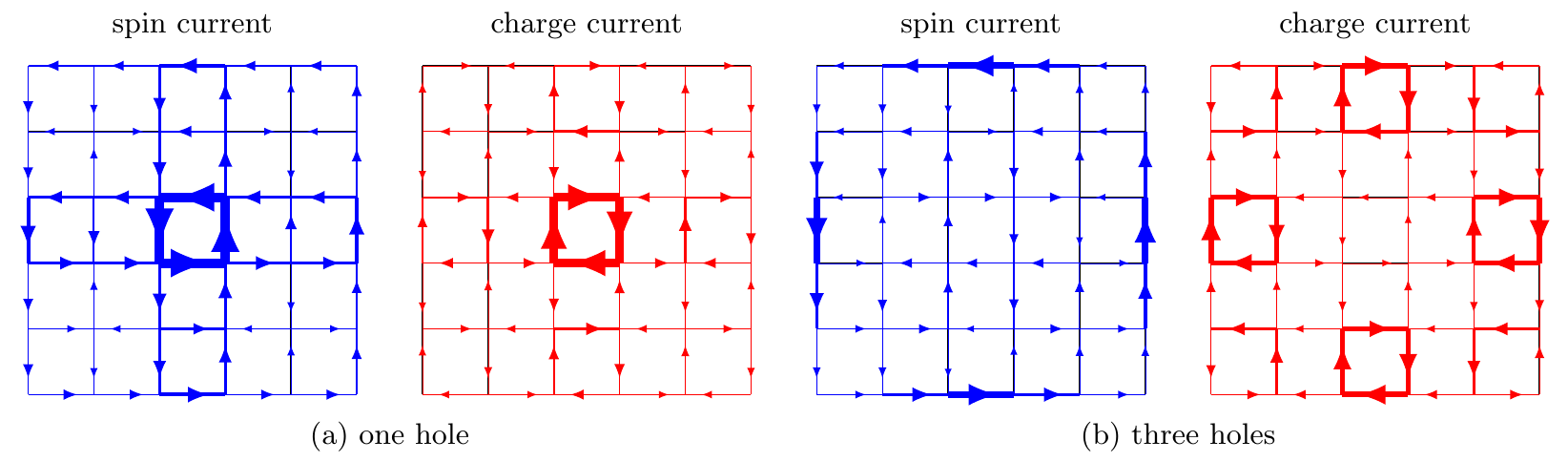}
    \caption{Spin currents are present in the one-hole and three-hole ground states which are doubly degenerate (for a fixed $S^z=1/2$) under the OBC with the angular momentum $L_z= \pm1  \mod 4$. But the spin current is absent in the two-hole ground state, which is non-degenerate with $L_z=2 \mod 4$ and $S=0$ consistent with a d-wave symmetry. Here $N=6\times{6}$ and $J/t=0.3$ with the data obtained by DMRG.}
    \label{fig:current_dmrg}
\end{figure*}

The amplitude of $I^x_s$ is non-universal and smoothly changes with $J/t$ as illustrated in Fig. \ref{fig:spin_current_trend}(a) for PBC, while the total momentum $\mathbf{K}^{0}$ jumps from  $(+\pi/2, +\pi/2)$ to $(+\pi/2, 0)$ around $J/t \simeq 2$. Actually the spin current $I^x_s$ and the ground state degeneracy simultaneously disappear at $J/t\simeq 7.0$ as indicated in the inset of Fig. \ref{fig:spin_current_trend} where $\mathbf{K}^{0}$ jumps from $(+\pi/2, 0)$ to $(0, 0)$.  Here one can clearly see that the novel ground state degeneracy and nonzero spin currents are concomitant. 
We also present larger system results as shown in Fig. \ref{fig:spin_current_trend}(b) for OBC. The finite spin current regime corresponds to $L_z=\pm 1$ with the critical transition points identified at $J_{c1}/t\simeq 0.28$ and $J_{c2}/t\simeq 1.3$ for $4\times 4$ and $J_{c1}/t\simeq 0.08$ and $J_{c2}/t\simeq 1.1$ for $6\times 6$, respectively. The critical points of $J_{c1}/t \simeq 0.02$ and $J_{c2}/t\simeq 1.1-1.2$ for $8\times 8$ are also determined by directly looking for the appearance/disappearance of the novel ground state degeneracy and nonzero spin currents. Clearly,  the spin current phase is robust for a wide
range of parameter $J/t$ for large systems.  The current patterns for $6\times 6$ and $8\times 8$ under the OBC identified by the DMRG calculation at $J/t=1/3$ can be found in Fig.~\ref{fig:current_dmrg} and Supplementary Material, respectively.

\begin{figure}[]
    \centering
    \includegraphics[width=.4\textwidth]{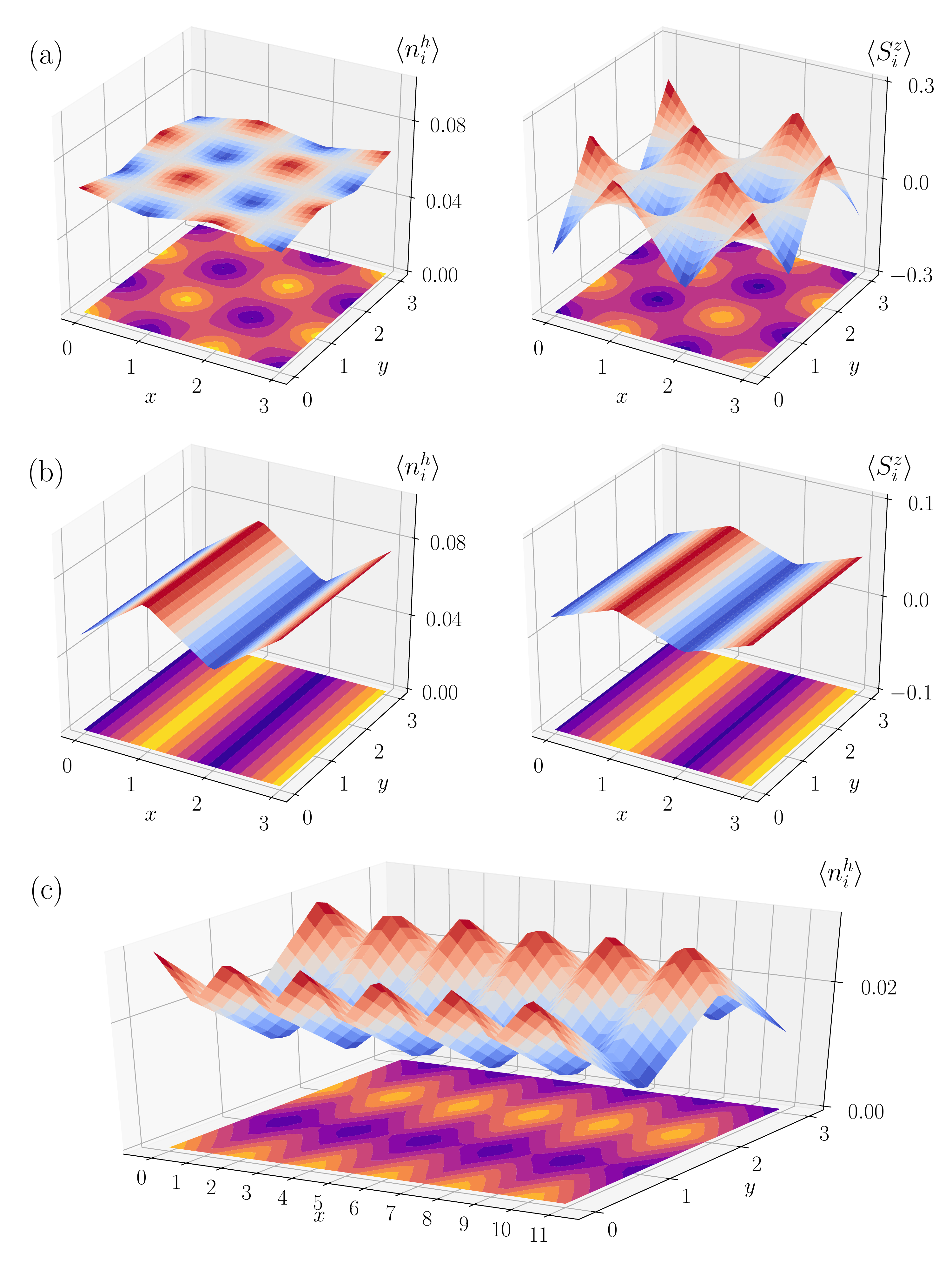}
    \caption{Charge and spin density modulations ($\langle{n}_{i}^{h}\rangle$ and $\langle{S}_{i}^{z}\rangle$, respectively) emerge in the degenerate ground states with (a) a net zero spin-current state; (b) a ``stripe'' state with zero net spin current only along the perpendicular direction. (c) Charge density wave obtained by DMRG. Here $J/t=0.3$ with $N=4\times 4$ in (a) and (b) and for $N=12\times 4$ in (c) under the PBC. }
    \label{fig:modulations}
\end{figure}

\emph{Charge/spin modulations.---}One may further construct a \emph{zero} or \emph{unidirectional} spin current state by proper superpositions of the current carrying states specified by the total momenta ${\bf K}^0=(\pm\pi/2, \pm\pi/2)$ discussed above. As illustrated by Fig. \ref{fig:modulations}(a) for the case of $N=4\times 4$, by an equal weight superposition of all four states, the new state exhibits both charge and spin modulations on top of a uniform background. Furthermore, a stripe-like charge/spin spatial distribution can be constructed in Fig. \ref{fig:modulations}(b) as a superposition of two degenerate ground states with vanishing spin currents perpendicular to the stripe direction (while the  quantized momentum remains along the stripe direction). Furthermore, an $N=12\times 4$ system calculated by DMRG shows the same four-fold degeneracy states at the same ${\bf K}^0$, whose real wave function is a zero-current state with the similar charge (spin) modulations as illustrated in Fig. \ref{fig:modulations}(c). Here the charge/spin modulations or nematicity as the ``incipient'' translational symmetry breaking \cite{1707.00068} may be viewed as many-body quantum interference states, which are ``dual'' to the degenerate spin-current-carrying ground states.

\emph{A few hole  cases.---}Now let us examine the case when more holes are injected into the Mott insulator. We have seen that there is a ground state degeneracy associated with nonzero spin currents in the one-hole case. Surprisingly, the whole ground-state degeneracy and neutral spin currents disappear simultaneously in the two-hole ground state.
In particular, the total angular momentum becomes $L_z=\pm 2 \mod 4$\cite{note2} at $J/t=0.3$, which is consistent with the d-wave symmetry of two hole pairing state (i.e., the wavefunction changes sign under a $\pi/2$ rotation). Note that previously a strong binding between the two holes has been indeed shown in the two- and four-leg ladders with $N=N_x\times 2$ and $ N=N_x\times 4$ by DMRG for  the same ratio of $J/t$ \cite{srep5419}.

However, once three holes are doped, the novel degeneracy and spin currents reemerge again in the ground states.  In Fig.~\ref{fig:current_dmrg}, both the spin and charge currents in the $N=6\times 6$ system are shown for (a) the one-hole case and (b) three-hole case as obtained by DMRG under the OBC.
We find that charge currents show different microscopic patterns with a staggered current loop pattern in the background \cite{PhysRevB.64.100506}\cite{PhysRevB.79.205115}\cite{PhysRevLett.90.186401}, and their amplitude distributions are correlated with the ones for the spin currents.
We always find the disappearance of the degeneracy and spin currents for the even-numbers of holes, while the irreducible double degeneracy (with a given $S^z\neq 0$) reemerges again when the number of doped hole is odd, where the nontrivial spin current persists up to an intermediate hole density for different system sizes and geometries as checked by DMRG.
For example, for an $N=6\times 6$, we find the same degeneracy with nonzero spin current pattern still present for the hole number equal to 5 (i.e., corresponding to the hole doping concentration $5/36\sim  14\% $).

\emph{Long-range entanglement due to a many-body Berry-like phase.---}The nonzero spin current is a demonstration of a Berry-like phase hidden in the background, which is nonlocally entangled with a doped hole as  clearly illustrated by, e.g., Fig.~\ref{fig:chain} (b). In the following, we  provide a theoretical understanding of its microscopic origin. It has been previously predicted that in the $t$-$J$ model a doped hole will generically pick up a Berry-like phase $\tau_c\equiv (-1)^{N^{\downarrow}_h(c)}$ after traversing the quantum spin background via a closed path $c$, which is known as the phase string effect \cite{PhysRevLett.77.5102}\cite{PhysRevB.55.3894}\cite{PhysRevB.77.155102}\cite{Zaanen2011}. Here $N^{\downarrow,\uparrow}_h(c)$ counts the total number of exchanges between the hole and $\downarrow$ ($\uparrow$) spins in the background with $\tau_c=e^{\pm i{\frac{\pi}{2}} [N^{\uparrow}_h(c)+N^{\downarrow}_h(c)]}e^{\mp i{\frac{\pi}{2}} [N^{\uparrow}_h(c)-N^{\downarrow}_h(c)]}$. 
It is distinguished \cite{ZZ2013} from the so-called $S^z$-string \cite{RevModPhys.78.17,Brinkman,Trugman1988,RevModPhys.66.763,Grusdt2017} as the \emph{transverse} component of the defect created by hole hopping.
We note that the first factor in $\tau_c$ will lead to $\mathbf{K}^{0}=(\pm\pi/2, \pm\pi/2)\neq 0$ while the second one will be responsible for generating the spin current as the residual fluctuations once the hopping $t$ becomes dominant locally. Indeed, by turning off the phase string $\tau_c$ with replacing the hopping term $H_t$ by $H_{\sigma\cdot{t}} = -t\sum_{\langle{ij}\rangle \sigma}\sigma(c_{i\sigma}^{\dagger}c_{j\sigma}+h.c.)$ in the so-called $\sigma\cdot$$t$-$J$ model \cite{ZZ2013}, all the above novel features disappear and we find unique ground state as confirmed by both ED and DMRG calculations. With $\tau_c=1$ and $\mathbf{K}^{0}=(0,0)$ or $(\pi,\pi)$, the ground state  reduces to a trivial ``quasiparticle'' state  without spin currents, and correspondingly it  becomes non-degenerate and uniform at a given $S^z=\pm 1/2$.

\emph{Summary.---}In this work, we have firmly established an important effect of the doped Mott insulator by ED and DMRG, which has been overlooked in the previous studies. Namely, a single hole or odd number of holes exhibits a composite structure by generating independent spin currents in the background. The latter should carry away a partial momentum or angular momentum. In the one-hole ground state, for example, the total momentum $\mathbf{K}^{0}=(\pm\pi/2, \pm\pi/2)$  has been previously well established \cite{RevModPhys.78.17}  \cite{RevModPhys.66.763,PhysRevB.52.R15711,SS1988,SCBA1,SCBA2,SCBA3,SCBA4} in the $t$-$J$ model and experimentally \cite{PhysRevLett.74.964}\cite{Ronning2067}. But the corresponding single-electron momentum distribution shows a much broadened feature (cf. Fig. S4 and the detailed discussion in the Supplementary Material). In particular, in contrast to a point-like quasiparticle without an internal degree of freedom, here the chirality of the spin current relative to the hole determines the sign of the total momentum/angular momentum and thus leads to a novel ground state degeneracy. The  doped hole is no longer a Landau's quasiparticle carrying the total momentum/angular momentum satisfying the one-to-one correspondence principle. On the other hand, the degenerate ground states with the charge and spin modulations can be reconstructed from the current-carrying states, with a period of doubled lattice constant in the one-hole case [cf. Fig. \ref{fig:modulations}(a)], which is consistent with the observation in the neighborhood of a trapped charge state by a defect in an undoped cuprate \cite{Ye2013}. Furthermore, the novel degeneracy, spin currents, and the charge/spin modulations all disappear in the case of even-number of holes, indicating that the spin currents must play an important role to facilitate pairing. Finally, if one makes an extrapolation to a finite hole density in the thermodynamic limit, the even-odd effect of doped holes could have a profound implication. If these holes are indeed paired up in the ground state to form a d-wave superconducting state, then a novel ``pseudogap'' phase may be conjectured at finite-temperature by the presence of a sufficient amount of unpaired single holes, where the finite spin and charge current loops as well as  charge/spin modulations or nematicity are expected to coexist. In particular, the charge modulation period would be changed, depending on a Fermi surface (pockets or arcs) emergent at finite doping as evolving from the four points at $\mathbf{K}^{0}$ in the one-hole case. These are open questions to be explored in future studies.

\begin{acknowledgements} Stimulating discussions with L. Balents, S. Chen, F.D.M. Haldane, J. Ho, S. Kivelson, J. Zaanen are acknowledged. This work is supported by Natural Science Foundation of China (Grant No. 11534007), MOST of China (Grant No. 2015CB921000, 2017YFA0302902).
Work by DNS is supported by  the DOE, through the Office of Basic Energy Sciences under the grant No. DE-FG02-06ER46305.

\end{acknowledgements}


\begin{thebibliography}{99}
\bibitem{note1} As shown in TABLE \ref{tab:degenerate_gs}, there are additional double degeneracies at $4\times 4$, which go away for larger systems \cite{PhysRevB.52.R15711}.
\bibitem{note2} The ground state with the angular momentum $L_z$ will change by a phase factor $e^{i\phi L_z}$ under a $\phi$ rotation, and on a square lattice with the rotational symmetry, $L_z$  is determined up to mod 4.
\end{thebibliography}

\renewcommand{\theequation}{S\arabic{equation}}
\setcounter{equation}{0}
\renewcommand{\thefigure}{S\arabic{figure}}
\setcounter{figure}{0}
\renewcommand{\bibnumfmt}[1]{[S#1]}

\clearpage
 \newpage
\onecolumngrid
\begin{center}
{\bf Hidden spin current in doped Mott antiferromagnets: Supplementary Material}
\end{center}

In this supplementary material,  we shall define the neutral spin current, backflow spin current, and charge current, respectively, and address the continuity conditions of the currents. The DMRG results of spin and charge currents for a $8\times8$ system doped by hole will be also shown.  Finally,  the violation of Landau's one-to-one correspondence conjecture will be discussed based on the momentum distribution function.

\section{Spin and charge currents}

Based on the $t$-$J$ model in Eq. (\ref{tj}), there are two globally conserved quantities, namely the hole number $N_h\equiv\sum_{i}(1-n_{i})=N-\sum_{i\sigma}c_{i\sigma}^{\dagger}c_{i\sigma}$ and the total magnetization $S_{tot}^{z}=\sum_{i}S_{i}^{z}$ as $[H, N_h] = 0$ and $[H, S_{tot}^{z}]=0$ in the restricted Hilbert space of $n_i\leq 1$. In the Heisenberg picture one has
\begin{equation}
    \frac{d[1-n_{i}(\tau)]}{d\tau}=\text{i}[H, 1-n_{i}]=-\text{i}(-t)\sum_{\langle jk \rangle, \sigma}\left[c_{j\sigma}^{\dagger}c_{k\sigma}+h.c., \sum_{\eta}c_{i\eta}^{\dagger}c_{i\eta}\right]\equiv\sum_{j=NN(i)}J_{ij}^{h},
    \label{eq:current_particle}
\end{equation}
in which the hole current is identified by
\begin{equation}\label {jh}
 J_{ij}^{h}=-\text{i}t\sum_{\sigma}(c_{i\sigma}^{\dagger}c_{j\sigma}-h.c.)~.
 \end{equation}

Similarly, for the local operator $S_{i}^{z}$
\begin{equation}    \label{sz}
    \begin{aligned}
        &\frac{dS_{i}^{z}(\tau)}{d\tau}=\text{i}[H_{t}, S_{i}^{z}]+\text{i}[H_{J}, S_{i}^{z}] \\
        &=\text{i}(-t)\sum_{jk, \sigma}\left[c_{j\sigma}^{\dagger}c_{k\sigma}+h.c., \frac{1}{2}\sum_{\eta}\eta{c}_{i\eta}^{\dagger}c_{i\eta}\right]+\text{i}J\sum_{jk}\left[\frac{1}{2}(S_{j}^{+}S_{k}^{-}+h.c.), \frac{1}{2}\sum_{\eta}\eta{c}_{i\eta}^{\dagger}c_{i\eta}\right] \\
        &\equiv \sum_{j=NN(i)}(J_{ij}^{b}+J_{ij}^{s}),
    \end{aligned}
\end{equation}
where the backflow current $J_{ij}^{b}$ associated with the hole hopping and the neutral spin current $J_{ij}^{s}$ in the spin background are respectively defined as follows:
\begin{subequations}
\begin{eqnarray}
    \\
    J_{ij}^{b}&=&\text{i}\frac{t}{2}\sum_{\sigma}\sigma(c_{i\sigma}^{\dagger}c_{j\sigma}-h.c.), \\
    J_{ij}^{s}&=&-\text{i}\frac{J}{2}\left (S_{i}^{+}S_{j}^{-}-h.c.\right)~.
\end{eqnarray}
\label{eq:spin currents}
\end{subequations}

\begin{figure}[]
    \centering
    \includegraphics[width=.9\textwidth]{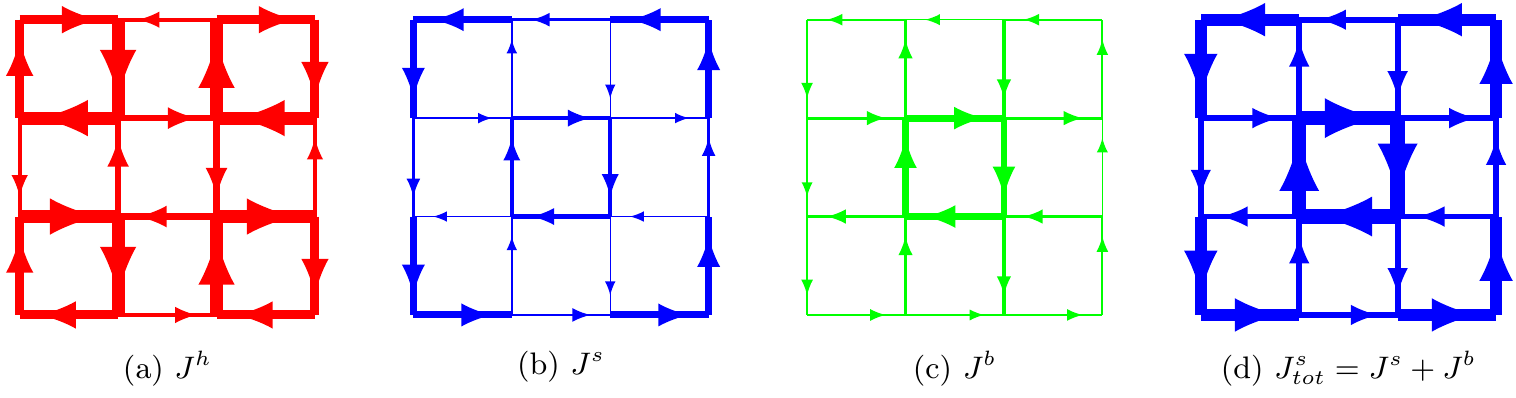}
    \caption{Currents of the one-hole ground state with $L_{z}=-1$ on a $4\times4$ lattice under the OBC with $J/t=0.3$.}
    \label{fig:full_currents_44}
\end{figure}

In the main text, for simplicity, in calculating the neutral spin current $J^s_{ij}$ we have set $J=1$ in the definition of $ J_{ij}^{s}$ in Eq. (\ref{sc}). Note that in order to have conserved currents, one has to include both $J^s$ and $J^b$ to restore the continuity condition. As illustrated in Fig. \ref{fig:full_currents_44}, we compute $J^h$, $J^s$, $J^b$, and $J_{\text{tot}}^s\equiv J^s+J^b$ in the $L_{z}=-1$ state of the $t$-$J$ model with $J/t=0.3$ and $N=4\times 4$ under the OBC. We have checked that the total spin currents in Fig. \ref{fig:full_currents_44}(d) does exactly satisfy the continuity condition.



\section{Neutral spin and charge currents at $N=8\times 8$ by DMRG}

\begin{figure}[]
    \centering
    \includegraphics[width=.6\textwidth]{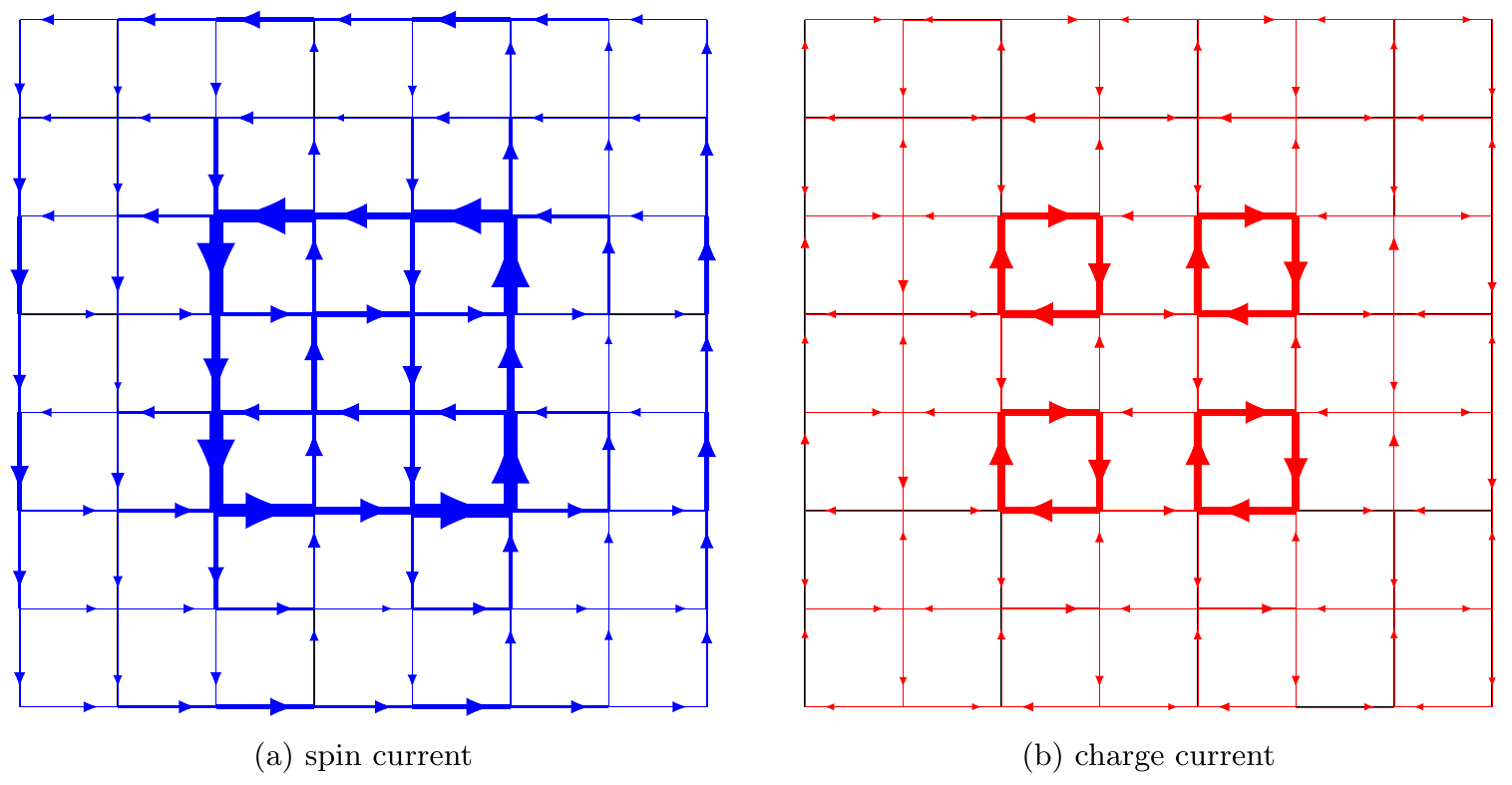}
    \caption{Spin and charge currents for the one-hole-doped $t$-$J$ model on a $8\times8$ lattice under the OBC with $J/t=0.3$. There are double degenerate ground states associated with $L^z=\pm 1$.}
    \label{fig:current_88}
\end{figure}

In the DMRG calculations,  it is usually difficult to directly select a translational invariant state with a given momentum quantum number due to the algorithm using local basis states \cite{PhysRevLett.69.2863}\cite{RevModPhys.77.259}. In our calculation,  we first calculate real wavefunctions which speed up the DMRG process.  However,  we can target different ground states and make superpositions
of these states to form  momentum or angular momentum eigenstates.   For an open system, we first obtain the lowest two energy eigenstates,  which are always degenerating
with each other for the one hole doped case with a suitable ratio of $J/t$.  The complex superpositions of these two ground
states ($(|\Psi_{01}\rangle\pm\text{i}|\Psi_{02}\rangle)/\sqrt{2}$)  will make up two angular momentum
eigenstates with $L_z=\pm 1$, respectively. We then can measure the spin and charge currents from one of these states, whose patterns
are shown in FIG. \ref{fig:current_88} for a lattice size $N=8\times8$ for the $t$-$J$ model at $J/t=0.3$. We see that the spin and charge currents in the ground state remain robust from $4\times 4$ to $8\times 8$, as well as $12\times 4$, which are tied up with the nontrivial exact ground state degeneracy at a fixed $S^z$. It is interesting to note that there is generally a staggered loop pattern in the background of the charge current shown in  FIG. \ref{fig:current_88}(b), which is consistent with that discussed in two-leg ladder systems \cite{PhysRevB.64.100506}\cite{PhysRevB.79.205115}\cite{PhysRevLett.90.186401}. Its details will be further discussed elsewhere.

\section{Momentum distribution: the breakdown of the one-to-one correspondence principle}

\begin{figure}[]
    \centering
    \includegraphics[width=.6\textwidth]{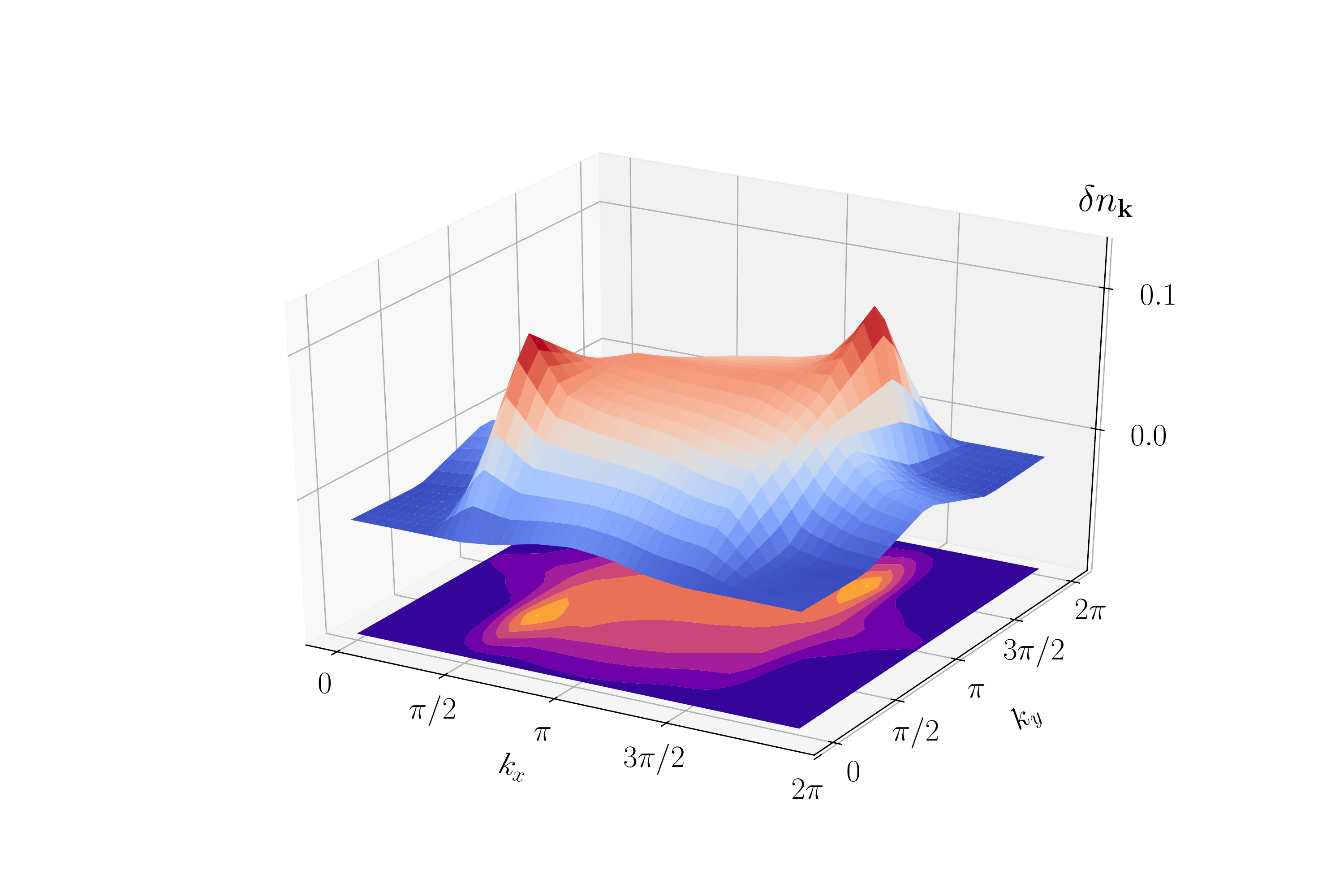}
    \caption{The \emph{change} of the electron momentum distribution, $\delta{n}(\mathbf{k})$, when one-hole is injected into the Mott insulator, is obtained by DMRG with lattice size $N=12\times4$ under the PBC with $J/t=0.3$. Note that the corresponding ground state shows charge modulation as given in Fig. \ref{fig:spin_current_trend} (c).}
    \label{fig:nk_3d}
\end{figure}

To further examine the physical implications of the presence of the neutral spin currents in the spin background,  we study the \emph{change} of the momentum distribution of the electrons upon doping:
\begin{equation}
    \delta{n}(\mathbf{k})\equiv n^e_0-n^e(\mathbf{k})=1-\sum_{\sigma}c_{\mathbf{k}\sigma}^{\dagger}c_{\mathbf{k}\sigma},
    \label{}
\end{equation}
where $n^e_0=1$ denotes the electron momentum distribution at half-filling (the Mott antiferromagnet). So  $\delta{n}(\mathbf{k})$ measures the \emph{change} of the electron momentum distribution upon one hole doping with $\sum_ {\mathbf{k}} \delta{n}(\mathbf{k})=1$.

Let us consider, as an example, an $N=12\times 4$ lattice with one doped hole under the PBC, which can be shown to have four-fold degenerate ground states at four total momenta $\mathbf{K}^{0}=(\pm \pi/2, \pm \pi/2)$ by our DMRG calculation. A real-wave-function ground state determined by DMRG exhibits the charge modulation as shown in  Fig. \ref{fig:modulations}, which is a superposition of the degenerate ground states of given $\mathbf{K}^{0}$'s. Correspondingly we examine the momentum distribution $\delta{n}(\mathbf{k})$  of such a ground state in the following.

\begin{figure}[]
    \centering
    \includegraphics[width=.6\textwidth]{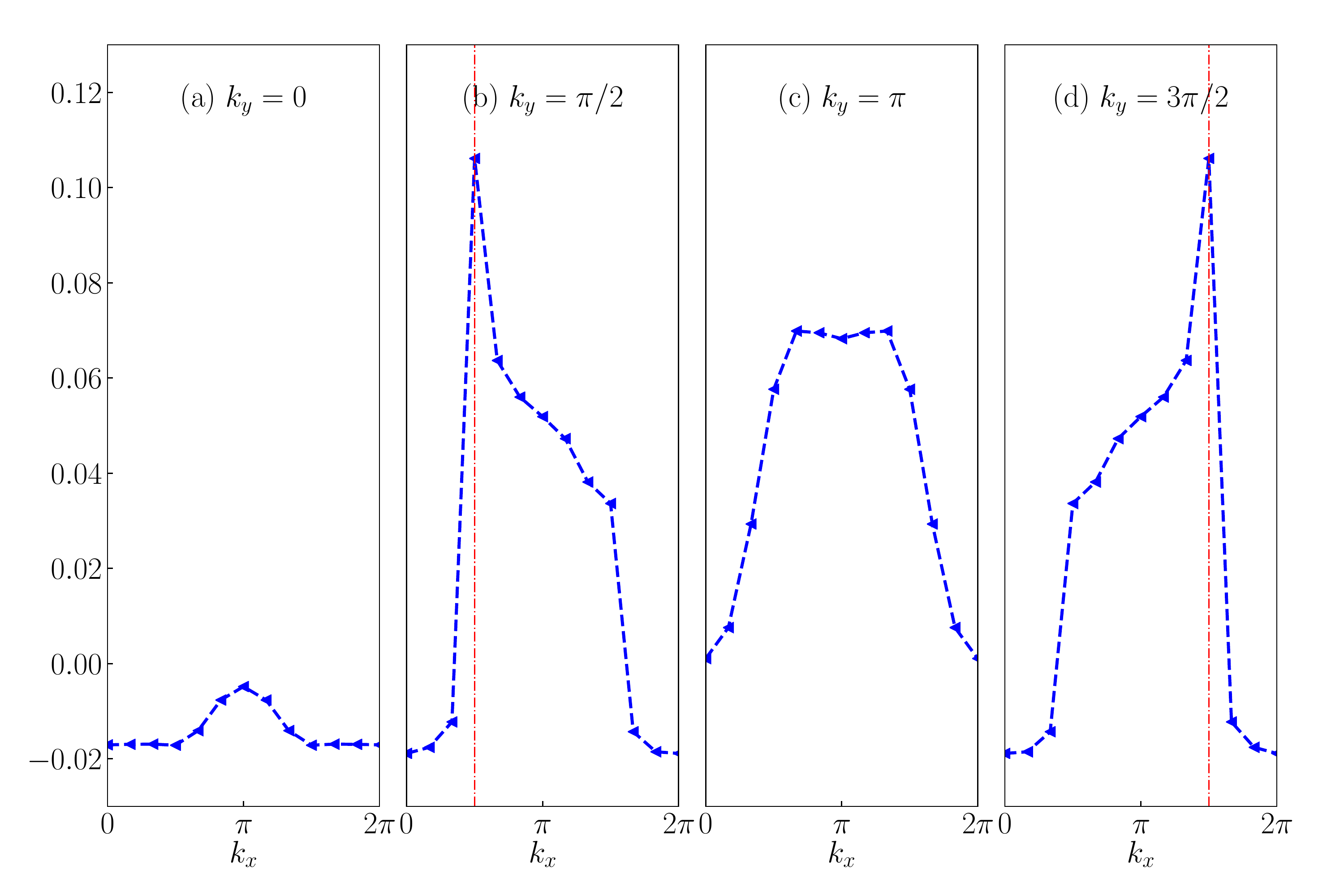}
    \caption{$\delta{n}(\mathbf{k})$ vs. $k_x$ at fixed $k_y$'s for the same ground state as in Fig. \ref{fig:nk_3d}. The vertical dashed lines mark the positions of the total momenta at $(\pi/2, \pi/2)$ and $(3\pi/2, 3\pi/2)$, by whose ground states the present degenerate state is superposed of.}
    \label{fig:nk}
\end{figure}

As shown by Fig. \ref{fig:nk_3d},  $\delta{n}(\mathbf{k})$ exhibits two major peaks located at $(\pi/2, \pi/2)$ and $(3\pi/2, 3\pi/2)$. The latter is equivalent to $(-\pi/2, -\pi/2)$ in the first Brillouin zone. However, $\delta{n}(\mathbf{k})$ clearly shows a continuum background, indicating that the individual electrons gain a broad range of momenta centered around the total  $\mathbf{K}^{0}$ upon one hole doping. Figure \ref{fig:nk} further illustrates the momentum distribution along the $k_x$-axis for given $k_y$'s. Both Figs. \ref{fig:nk_3d} and \ref{fig:nk} directly indicate that the total momentum is no longer solely carried by a single charge carrier or ``quasiparticle''. In other words, Landau's one-to-one correspondence principle, which is the basis for a Fermi liquid, is violated here.

The persistent spin currents in the spin background provide a microscopic mechanism for such a breakdown of the one-to-one correspondence. Indeed, the total momentum is associated with the translational symmetry of the \emph{whole} many-body system, which includes both the doped hole and the background spins. On the other hand, the concomitant spin currents will carry away partial momentum and the momentum transfer between the two degrees of freedom is generally present. In other words, the hole is moving in a quantum spin background which is not translational invariant as far as the doped charge is concerned.  As a matter of fact, it has been shown in Fig. \ref{fig:spin_current_trend}  that the strength of the spin currents is non-universal and smoothly changes with the coupling ratio $J/t$. As the consequence, it implies that the adiabatic continuity should no longer be valid here even though $\mathbf{K}^{0}$ is still well defined. A in-depth analysis of breakdown of the one-to-one correspondence for the two-leg ladder Mott insulators doped by a hole has been recently given in Ref. \onlinecite{1707.00068}.

\end{document}